\documentclass[twocolumn,showpacs,preprintnumbers,superscriptaddress]{revtex4-1}

\usepackage{amsfonts}
\usepackage{graphicx}
\usepackage{amsmath}
\usepackage{amssymb}
\usepackage{latexsym}
\usepackage{psfrag}
\usepackage[bookmarks, colorlinks=true, plainpages = true, linkcolor = blue, citecolor = red, urlcolor = blue, filecolor = blue]{hyperref}

\begin{document}
\title{Identification of the Molecule-Metal Bonding Geometries of Molecular Nanowires}

\author{Firuz Demir} 
\affiliation{Department of Physics, Simon Fraser
University, Burnaby, British Columbia, Canada V5A 1S6}

\author{George Kirczenow} 
\affiliation{Department of Physics, Simon Fraser
University, Burnaby, British Columbia, Canada V5A 1S6}

\date{\today}

\begin{abstract}\noindent
Molecular nanowires in which a single molecule bonds chemically to two metal 
electrodes and forms a stable electrically conducting bridge between them
have been studied intensively for more than a decade. However the 
experimental determination of the bonding geometry between the molecule and
electrodes has remained elusive. Here we demonstrate by means of {\em ab initio} 
calculations that inelastic tunneling spectroscopy (IETS) can determine these
geometries. We identify the bonding
geometries at the gold-sulfur interfaces of 
propanedithiolate molecules bridging gold electrodes that give rise to the specific IETS
signatures that were observed in recent experiments.

\end{abstract}

 
\maketitle

Molecular nanowires in which a single organic molecule bonds chemically to two
metal electrodes and forms a stable electrically conducting bridge between
them have attracted a great deal of attention\cite{review2010}  
because of their fundamental interest and potential applications as 
single-molecule nanoelectronic devices.
However, a single molecule located between two
electrodes is not accessible to scanning microprobes
that can measure atomic scale structure. Thus  definitive  determination of the
bonding geometries at the molecule-electrode interfaces of single-molecule
nanowires continues to be elusive despite being critically important for
understanding electrical conduction through molecular wires quantitatively, 
and for gaining control over their structures for device
applications. 

In the case of
single-molecule nanowires bridging gold electrodes and thiol bonded to them,
possible bonding geometries include
those in which a sulfur atom of the molecule is located at a {\em top} site
over a particular gold surface atom or over a {\em bridge} site between two
gold atoms or over a {\em hollow} site between three gold
atoms.\cite{review2010} 
However, there have
been no direct experiments determining which (if any) of these possibilities
are actually realized. For molecules amine-linked to gold electrodes 
it has been argued that top-site bonding is the most probable\cite{Hybertsen}, but {\em direct} 
experimental evidence of this has also been lacking.   

The excitation of molecular vibrations (phonons) by electrons
passing through single-molecule nanowires gives rise to conductance steps in
the low temperature current-voltage characteristics of the nanowires.
Inelastic tunneling spectroscopy (IETS) experiments  have detected these steps
and measured the energies of the emitted phonons
\cite{IETSexperiments}. 
These experiments proved that particular molecular species are involved in electrical
conduction through metal-molecule-metal junctions. 
Density functional theory (DFT) based simulations \cite{Pecchia2004, 
TroisiRatner05, IETSDFT}
have accounted for many features of the IETS data.
However, the possibility that IETS might identify the {\em bonding geometries} at the
molecule-metal {\em interfaces} and thus resolve the long standing problem of
the molecule-electrode structure has not been investigated in
the literature. We explore it
in this paper and identify for the first time the
molecule-contact bonding geometries that were realized in experiments on an
organic molecule bridging metal contacts. We consider one of the
simplest organic molecules, 1,3-propanedithiolate (PDT), bridging gold
electrodes for which detailed experimental IETS data is 
available.\cite{HihathArroyoRubio-BollingerTao08}

We focus on inelastic tunneling processes that are sensitive to
the structure of the gold-sulfur interfaces, i.e., those that involve 
excitation of vibrational modes with strong amplitudes on the
sulfur atoms. Therefore it is necessary to calculate accurate equilibrium
geometries and also  the frequencies and atomic displacements from equilibrium
for the vibrational modes of the {\em whole} system, including
both the molecule and the gold electrodes. We do this by performing {\em ab
initio} DFT calculations\cite{Gaussian} for {\em extended}
molecules consisting of a PDT molecule and two finite clusters of
gold atoms that the molecule connects, relaxing this entire structure.  
By carrying out systematic DFT studies of extended
molecules
with gold clusters of different sizes (up to 13 gold atoms per cluster) we
establish that our conclusions are independent of the cluster size for the larger
clusters that we study and thus are applicable to molecules
bridging the nanoscale tips of experimentally realized macroscopic gold electrodes.

We found extended molecules whose sulfur atoms bond to the
gold clusters over bridge sites to have lower energies than those with top
site bonding by at least 0.76 eV. Extended molecules for which DFT geometry
relaxations were started with the sulfur atoms over hollow sites on
the surfaces of close packed gold clusters invariably relaxed to bridge
bonding site geometries.  While we found it possible to generate examples of
relaxed extended molecules with each sulfur atom bonding to three gold atoms
(i.e., hollow site bonding) the structures of the gold clusters near these
bonding sites were much more open than that of a fcc gold crystal. The
energies of these structures were also higher than those of either bridge or
top site bonded extended molecules. Because of the much greater fragility 
and higher energies of hollow site bonded structures relative to bridge and 
top site bonding we will focus primarily on bridge and top site
bonding, but will revisit hollow site bonding near the end
of this paper.

After identifying the normal modes of the extended molecule that have the
largest vibrational amplitudes on the sulfur atoms and calculating the
frequencies of those modes as is outlined above, we determine which of these
modes has the strongest IETS intensity, i.e, gives rise to the largest
conductance step height as the bias voltage applied across the extended
molecule is varied.

We calculate the IETS intensities perturbatively 
in the spirit of an approach proposed by 
Troisi {\em et al.}\cite{TroisiRatnerNitzan03} who transformed 
the problem of calculating IETS
intensities into an {\em elastic} scattering problem. However, we formulate our
IETS intensities explicitly in terms of elastic electron transmission
amplitudes  $ t_{ji}^{el}$ through the molecular wire. We find 
the height $\delta g_{\alpha}$ of the
conductance step  due to the emission of phonons of
vibrational mode $\alpha$ to be
\begin{equation} 
\delta g_{\alpha}=  \frac{e^2}{2\pi \omega_{\alpha}} \lim_{A\to0}\sum_{ij} 
\frac{v_j}{v_i }   
\Big\vert \frac{ t_{ji}^{el}(\{A{\bf d}_{{n}\alpha}\})-t_{ji}^{el}
(\{{\bf 0}\})} { A }
 \Big\vert ^2 
\label{intensityresult} 
\end{equation}
at low temperatures. Here $t_{ji}^{el}(\{{\bf 0}\})$ is the elastic 
electron transmission amplitude through the molecular wire in its
equilibrium geometry from state $i$ with velocity $v_i$ in the electron source
to state $j$ with velocity $v_j$ in the electron drain.  
${\bf d}_{{n}\alpha}$ represents the displacements from their equilibrium 
positions of the atoms $n$ of the extended molecule in normal mode 
$\alpha$ normalized so that $\sum_{n}  m_n  {\bf d}^*_{{n}\alpha'}\cdot
{\bf d}_{{n}\alpha} = \delta_{\alpha',\alpha}$ where $m_n$ is the mass of 
atom $n$. $\omega_{\alpha}$ is the frequency of mode $\alpha$. 
$t_{ji}^{el}(\{A{\bf d}_{{n}\alpha}\})$ is the elastic electron 
transmission amplitude through the molecular wire with each atom $n$
displaced from its equilibrium position by $A{\bf d}_{{n}\alpha}$ where $A$ is
a small parameter.
Our formal derivation of Eq.~(1) will be presented
elsewhere; here we point out that  Eq.~(1) is plausible
intuitively in a similar way to the result of 
Troisi {\em et al.}\cite{TroisiRatnerNitzan03}~: Eq.~(1) 
states that in the leading order of perturbation theory the scattering amplitude for {\em
inelastic} transmission of an electron through the molecular wire is
proportional to the change in the {\em elastic} amplitude for
transmission through the wire if its atoms are displaced from their equilibrium
positions as they are when vibrational mode $\alpha$ is excited.

In our evaluation of $ t_{ji}^{el}$
in Eq.~(1) the coupling of the extended molecule to
the macroscopic electron reservoirs was treated as in previous 
work\cite{silicon, DalgleishKirczenow,
Cardamone08} by attaching a large number of
semi-infinite quasi-one-dimensional ideal leads to the valence orbitals of the
outer gold atoms of the extended molecule. The transmission amplitudes
$t_{ji}^{el}$ were then found by solving the Lippmann-Schwinger equation
\begin{equation}\label{Lippmann-Schwinger} 
\vert \Psi \rangle =  \vert \Phi_0 \rangle +
G_0(E) W  \vert \Psi \rangle 
\end{equation}
where $ \vert \Phi_0 \rangle$ is an electron eigenstate of 
an ideal left lead that is decoupled from the extended
molecule,  $G_0(E)$ is the Green's function for the decoupled system of the ideal leads 
and the extended molecule, $W$
is the coupling between the extended molecule and leads, and $\vert \Psi
\rangle$ is the scattering eigenstate of the complete coupled system. In
evaluating $G_0(E)$ semi-empirical extended H\"{u}ckel  theory\cite{yaehmop,ammeter78}
was used to model the electronic structure of the extended molecule. Previous
calculations based on extended H\"{u}ckel  theory have yielded elastic
tunneling conductances in agreement with experiment for molecules
thiol bonded to gold electrodes\cite{DattaEmberly,
Kushmerick02, Cardamone08} and have accounted for 
transport phenomena observed in molecular arrays
on silicon\cite{silicon} as well as
electroluminescence data\cite{Buker08}, current-voltage
characteristics\cite{Buker08} and STM images\cite{Buker05} of molecules on
complex substrates.
 
We calculated the zero bias tunneling conductances for gold-PDT-gold molecular
wires from the Landauer formula $g = g_0 \sum_{ij}    \vert
t_{ji}^{el}(\{{\bf 0}\}) \vert ^2 v_j/v_i$ with $g_0=2e^2/h$, evaluating the
elastic transmission amplitudes $t_{ji}^{el}(\{{\bf 0}\})$ as described above
and found values in the range 
$g=$0.0012-0.0014$g_0$ and 0.0012-0.0020$g_0$ for
top site and bridge site bonded molecules, respectively. The degree of
agreement between these theoretical values and the experimental
value\cite{HihathArroyoRubio-BollingerTao08} $0.006 \pm 0.002 g_0$ is
typical of that in the literature\cite{Lindsay07} on molecules thiol
bonded to gold electrodes; as in previous studies\cite{Lindsay07} comparing
the elastic conductance calculations with experiment  does not reveal which
bonding geometry was realized experimentally.

We then calculated the vibrational modes, their frequencies and IETS
intensities as described above for many gold-PDT-gold extended molecules with
various gold-sulfur bonding geometries and compared our results with the
experimental data of 
Hihath \emph{et al.}\cite{HihathArroyoRubio-BollingerTao08}. 
Our calculations showed that the modes with
strong amplitudes of vibration on the sulfur atoms that have the largest IETS
intensities fall within the phonon energy range of a prominent feature of the
experimental IETS phonon histogram\cite{HihathArroyoRubio-BollingerTao08} 
that extends from $39$ to $52$~meV.

\begin{figure}[here] 
\includegraphics[width=0.99\linewidth]{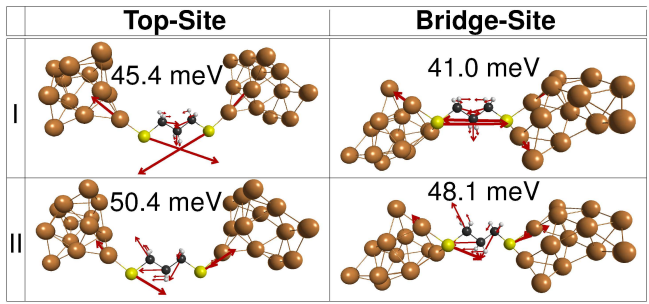} 
\caption{ Calculated vibrational modes in  
phonon energy range from $39$ to $52$~meV for trans-PDT 
bridging gold nano-clusters with sulfur atoms bonded to gold in
top-site and bridge-site geometries. Red arrows show (arbitrarily normalized) 
atomic displacements.
Mode I has the stronger IETS
intensity.\cite{Macmolplt}
}
\label{Figure1} 
\end{figure}

\begin{figure}[here] 
\includegraphics[width=1.0\linewidth]{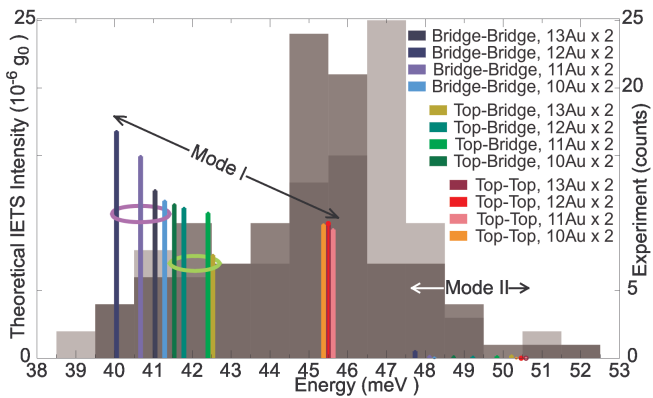}   
\caption{ Calculated phonon energies and 
IETS intensities for trans-PDT molecules linking pairs of gold 
clusters with between 10 and 13 Au atoms in each cluster.
Results are shown for both sulfur atoms bonding to the gold 
in top-site and bridge-site geometries
and for top site bonding to one gold cluster and bridge site
bonding to the other. The two ellipses enclose the calculated type I mode IETS spectra
for pure bridge and mixed top-bridge bonding geometries respectively for
extended molecules with gold clusters containing various numbers of gold atoms. 
The experimental IETS phonon mode
histogram of 
Hihath \emph{et al.}\cite {HihathArroyoRubio-BollingerTao08} is
shown in (darker, lighter) grey for (positive, negative) bias voltages.
}
\label{Figure2} 
\end{figure}

We show the vibrational modes that we find in this energy range for
representative top site and bridge site bonding geometries in
Fig. \ref{Figure1}. The calculated IETS spectra (phonon energies and IETS
intensities) in the same energy range are shown  in Fig.
\ref{Figure2} for  several extended
molecules with gold clusters of various sizes  together with the experimental
IETS phonon mode histogram.\cite {HihathArroyoRubio-BollingerTao08} 

The mode with the strongest calculated IETS intensities in Fig.~\ref{Figure2} is mode I
shown in the top row of Fig.~\ref{Figure1}. In this mode the sulfur atoms have
the strongest vibrational amplitudes and move in antiphase, approximately
along the axis of the molecule. The other mode in this energy range is mode
II shown in the lower row of Fig.~\ref{Figure1}. It is similar to mode I
except that the sulfur atoms move in phase with each other. As is seen in 
Fig. \ref{Figure2} the calculated IETS intensities for mode II are much weaker
than those for mode I for both top and bridge site bonding.
%
%
%
%
%
%
%
%
The reason for this difference between the IETS intensities of modes I and II
can be understood intuitively by considering the nature of the motion in
relation to  Eq.~(1): Since in mode I the two sulfur atoms
move in antiphase the gold-sulfur distances for both sulfur atoms either
increase or decrease {\em together} as the extended molecule vibrates. These 
distances can be regarded as the widths of tunnel barriers between the
molecule and the two gold electrodes. Thus the
motions of the two sulfur atoms act {\em in concert} to widen or narrow
{\em both} tunnel barriers together and therefore to weaken or
strengthen 
the electron transmission amplitude through the molecular wire. 
Thus the magnitude of the difference between the elastic
electron transmission amplitudes through the molecular wire in its equilibrium
 and vibrating geometries 
{\normalsize $  \big\vert t_{ji}^{el}(\{A{\bf
d}_{{n}\alpha}\})-t_{ji}^{el}(\{{\bf 0}\}) \big\vert$} in Eq.~(1) is enhanced. 
By contrast in mode II when the gold-sulfur
distance for one sulfur atom increases the gold-sulfur distance for the  other
 sulfur atom decreases. Thus the effects of the motions of the two sulfur
atoms on the elastic transmission amplitude through the molecular wire tend to
cancel. Thus 
{\normalsize $  \big\vert t_{ji}^{el}(\{A{\bf
d}_{{n}\alpha}\})-t_{ji}^{el}(\{{\bf 0}\}) \big\vert$ }  in Eq.~(1) 
is smaller for mode II than mode I and therefore the
IETS intensity $\delta g_{\alpha}$ for mode II is {\em much} smaller as 
is seen in Fig. \ref{Figure2}.

Comparison of the calculated IETS spectra in  Fig. \ref{Figure2} with the
experimental phonon mode histogram\cite{HihathArroyoRubio-BollingerTao08}
indicates that vibrational mode I contributed most of the counts recorded
in the histogram in the energy range shown. This is consistent with the
fact that the calculated IETS intensities for mode I are much stronger than
those for mode II and therefore mode I should be more readily detected in
experimental IETS measurements.

The theoretical results for the dominant IETS mode I in Fig. \ref{Figure2} reveal that the
prominent features of the histogram\cite{HihathArroyoRubio-BollingerTao08} 
can be explained as arising from PDT molecules that bond to the gold electrodes in {\em different} 
ways: The main peak in the experimental 
histogram\cite{HihathArroyoRubio-BollingerTao08} that is 
centered at $\sim$46~meV matches our theoretical result for trans-PDT molecules that bond to 
both gold electrodes in the
top site geometry. The weaker peak centered near $42$~meV 
matches our predictions for molecules that bond to one gold electrode in the 
top site geometry and to the other electrode in the bridge site geometry. The
shoulder of the experimental 
histogram\cite{HihathArroyoRubio-BollingerTao08}  at lower phonon energies 
between $39.5$~meV and $41.5$~meV corresponds to our results for molecules 
bonding to both electrodes
in the bridge site geometry.

Notice that our theoretical results for phonon mode I of PDT molecules
bonded to both electrodes in the top site geometry (the feature near 
45.5 meV in the theoretical spectra in Fig. \ref{Figure2}) are very
well converged with respect to increasing gold cluster size: Both the calculated phonon
energies and the IETS intensities are almost independent of the gold cluster
size in the size range shown (10-13 gold atoms per cluster). The calculated phonon
energy of this dominant mode matches  the phonon energy of
the main peak of the experimental phonon mode histogram in Fig. \ref{Figure2} very well.
It is also well separated from the calculated phonon energies of the dominant mode I
for the pure bridge and mixed bridge-top site bonded geometries.
As a check we calculated the vibrational mode energies 
and IETS intensities for a few examples of extended molecules using a different 
density functional\cite{Gaussian} and found similar results. 
Thus our results identify 
{\em unambiguously} those specific
realizations of the molecular wire that gave rise to the counts within the main peak 
of the experimental
histogram\cite{HihathArroyoRubio-BollingerTao08}  reproduced in Fig. \ref{Figure2} 
as being those in which both sulfur atoms bonded to the gold
electrodes in the top site geometry. 
The calculated mode I phonon energies in Fig. \ref{Figure2}
for pure bridge site-bonded and mixed bridge-top site-bonded molecular wires 
show more variation with gold cluster size than do the calculated mode I phonon energies
for molecules in the pure top site-bonded geometry. However, the ranges in
which the calculated energies of the mode I phonons 
for pure bridge site-bonded and mixed bridge/top site-bonded molecular wires occur 
do not overlap. Also, the calculated phonon energies for these modes and bonding geometries do not show
any systematic trend towards higher or lower values with increasing
cluster size. Thus it is plausible that the weaker peak centered near $42$~meV 
in the experimental histogram is due to mixed bridge/top site-bonded molecular wires
and the lower energy shoulder between 39.5 meV and 41.5 meV in the histogram 
is due to pure bridge site-bonded wires.

As the gold-PDT-gold junction was stretched in the 
experiment\cite{HihathArroyoRubio-BollingerTao08} the energy of the
prominent phonon mode in the IETS spectrum 
was observed to switch from $\sim$42 meV to $\sim$46 meV.
It was conjectured\cite{HihathArroyoRubio-BollingerTao08} that this switch
may be due to a change in the contact
configuration or between gauche and trans molecular geometries, 
but no evidence supporting either possibility was offered
and the contact configurations involved were not 
identified.\cite{HihathArroyoRubio-BollingerTao08} Here we point out
that trans-PDT molecules switching from the mixed top-bridge site bonding
(calculated phonon energy $\sim$42 meV)
to the pure top site bonding geometry (calculated phonon energy $\sim$45.5 meV)
accounts for this observed transition. 
Our calculated distance between the gold clusters of the extended molecule is larger for
pure top site bonding than for mixed top-bridge bonding. This is consistent with 
the transition from mixed top-bridge to pure top site bonding occurring
as the molecular junction is stretched.
For longer chain alkanedithiolates, it has been conjectured\cite{Li2006} that switching from mixed hollow site-top site  bonding, to top site bonding at both gold electrodes
may occur. This conjecture does not account for the observed switch
from the $\sim$42 meV mode to the $\sim$46 meV mode in the gold-PDT-gold junctions: As noted above, we find hollow-site bonding to be much more fragile (and thus less likely to be realized) than bridge site bonding. We also find the energy of the mode of the mixed hollow-top structure with the strongest IETS intensity to match {\em neither} the $\sim$42 meV nor the $\sim$46 meV phonon mode. 
The theoretical
results presented above are for {\em trans}-PDT molecules. We have also studied many gold-PDT-gold structures with molecular gauche defects and find that such structures also do not account for the 
observed\cite{HihathArroyoRubio-BollingerTao08} switching 
from the $\sim$42 meV phonon mode to the $\sim$46 meV phonon mode.
 
In conclusion: We have shown inelastic tunneling spectroscopy to be able to
distinguish between different bonding geometries of the molecule
and metal contacts in single-molecule molecular wires, an important and
previously elusive goal in the field of single-molecule nanoelectronics.  
We have definitively identified particular realizations of
gold-propanedithiolate-gold molecular wires in a recent 
experiment\cite{HihathArroyoRubio-BollingerTao08} in which 
the molecule bonded to a single
gold atom of each electrode. 
The success of our approach rests on the fact
that {\em ab initio} density functional theory calculations of vibrational modes and
their frequencies are known to be accurate because in the Born-Oppenheimer
approximation they are electronic {\em ground state total energy}
calculations.\cite{review2010, TroisiRatner05} We rely on transport calculations only for 
the identification of the phonon mode in a
particular frequency range that has the largest IETS intensity, and our
identification of this mode is also supported by physical  reasoning.

This research was supported by CIFAR, NSERC and Westgrid. 
We thank J. Hihath, N. J. Tao, E. Emberly, N. R. Branda, V. E. Williams and A. Saffarzadeh 
for helpful discussions and 
 J. Hihath and N. J. Tao for providing to us their data from 
Ref.\citenum{HihathArroyoRubio-BollingerTao08} 
in digital form.

\end{document}